\documentclass[12pt]{revtex4-1}
\usepackage{amsmath,graphicx,hyperref,bm}
\usepackage{amssymb}
\usepackage[font={small},flushleft,indent]{caption}
\usepackage{booktabs}
\begin{document}
\title{Redshift drift in uniformly accelerated reference frame\thanks{Supported by National Natural Science Foundation of China (11675182, 11690022)}}

	\author{Zhe Chang,}
	%\author[1]{and Qing-Hua Zhu,\note{Corresponding author.}}
	\author{Qing-Hua Zhu}
	\email{zhuqh@ihep.ac.cn}
	
	\affiliation{Institute of High Energy Physics, Chinese Academy of Sciences, Beijing 100049, China}
	\affiliation{University of Chinese Academy of Sciences, Beijing 100049, China}

\begin{abstract}
We construct  an alternative uniformly accelerated reference frame based on the 3+1 formalism in adapted coordinates. In this frame, time-dependent redshift drift exists between co-moving observers, which differs from that in Rindler coordinates. This phenomenon can be tested in laboratory and improve our understanding of non-inertial frames.
\end{abstract}

  \maketitle
\section{Introduction}

Owing to the special relativity, inertial frames are well tested and understood. The principle of relativity indicates that physical equations remain the same in all inertial frames. For non-inertial frames, there is a general principle of relativity; namely, the physical equations remain the same in arbitrary reference frames. This principle should involve all non-inertial reference frames. However, even uniformly accelerated reference frames are not yet understood well \cite{marzlin_what_1996}. And different uniformly accelerated reference frames are set up from different points of views \cite{lass_accelerating_1963,rindler_kruskal_1966,huang_new_2006,alba_generalized_2007,felix_da_silva_space_2007}.

Propagation of light in non-inertial frames provides a way for testing relativity in non-inertial reference frames. The Sagnac effect states that in a rotating reference frame, counter-propagating rays that propagate around a closed path would take different time intervals \cite{post_sagnac_1967,wang_generalized_2004}. This can be described by the Born metric, known as a relativistic effect % and would be interesting to explore that in general relativity
\cite{landau_classical_1980, ashtekar_sagnac_1975,benedetto_general_2019}. Likewise, does a similar effect exist in uniformly accelerated frames? As we know in the view of inertial observers, accelerated detectors would observe a time-dependent redshift of light from a co-moving source. %There is velocity difference, when the light is emitted and received. 
Could the redshift drift be observed  in uniformly accelerated reference frames as a relativistic effect? %And how the co-moving object is ?

In order to answer these questions, a metric of uniformly accelerated reference frames should be given explicitly.  Rindler coordinates \cite{rindler_kruskal_1966}, also named as Møller coordinates or Lass coordinates \cite{lass_accelerating_1963}, are generally used in uniformly accelerated reference frames. %\cite{davies_scalar_1975,fulling_s._a._radiation_1976,unruh_notes_1976}.
As rigid coordinates, they do not present a redshift drift. This seems not
consistent with the observations from inertial observers. Huang \cite{huang_uniformly_2008} suggested that the redshifts without a drift in Rindler coordinates should be attributed to the norms of four-accelerations, which are not the same for all co-moving observers. Besides, Minser, Thorne, and Wheeler (MTW) \cite{misner_gravitation_1973} derived  Møller coordinates with the hypothesis of locality. This indicates that Rindler coordinates are in fact local frames. The redshift  drift  might be a higher order effect. All these considerations motivated us to construct an alternative uniformly accelerated reference frame that is different from Rindler coordinates and  the local frame \cite{misner_gravitation_1973,ni_inertial_1978,mashhoon_hypothesis_1990,marzlin_what_1996,alba_generalized_2007,gourgoulhon_auth._special_2013}.

In this study, we investigate an adapted coordinate in which all co-moving observers have the same norms of four-accelerations. Explicit metric and coordinate transformation are obtained. 
Moreover, the redshifts for co-moving observers in the new uniformly accelerated reference frames are
calculated. Using the new proposed uniformly accelerated
frames, we investigate a possible Unruh effect and show a non-thermal
distribution perceived by the uniformly accelerated observers in Minkowski vacuum.

This paper is organized as follows. In section \ref{II}, we review the
redshift between co-moving objects in a non-relativistic approximation and in Rindler coordinates. We find that there is a redshift drift  in the non-relativistic approximation, whereas this is not in the Rindler coordinates.
In section \ref{III}, we present the construction of a uniformly accelerated reference frame
as well as its features. In section \ref{IV}, we provide
explicit metrics of the accelerated frames. The redshift drift
and the possible Unruh effect in the accelerated frames are studied.
Finally, conclusions and discussions are summarized in section \ref{VI}. Throughout, we use the
convention that $c = 1$.

\section{Redshift drift and uniformly accelerated reference frames}\label{II}

\begin{figure}[h]
\includegraphics[width=0.6\linewidth]{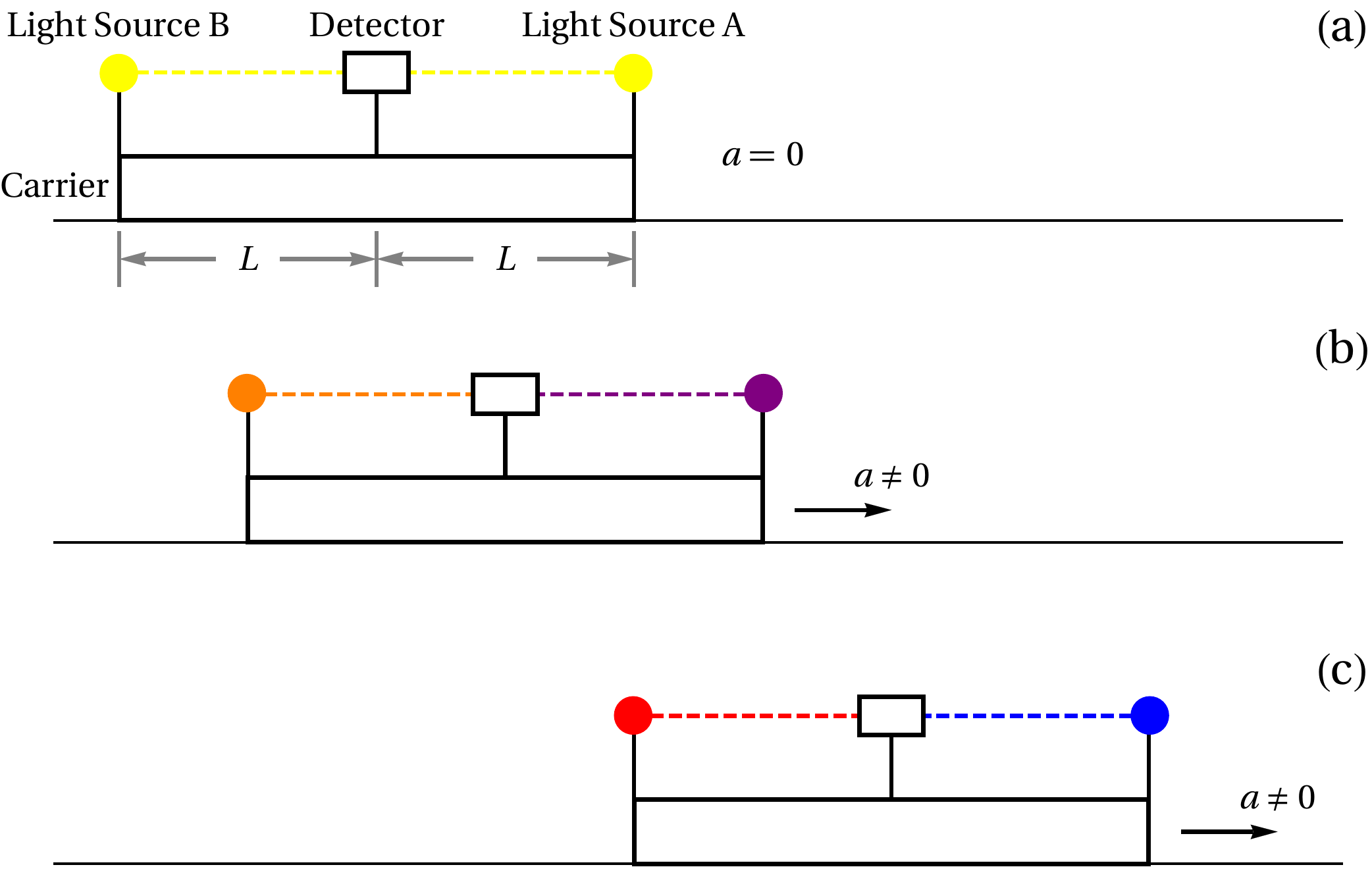}
\caption{Schematic of redshift drift in uniformly accelerated reference frames. Light sources A and B and a detector are fixed on carrier. Panel (a): No redshift is observed by detector when carrier undergoes inertial motion.   Panel (b): When carrier moves to right with constant acceleration $a_0$, detector will observe redshift from source B and blueshift from source A. Panel (c): Observed redshift and blueshift will drift with time, if carrier remains in uniformly accelerated motion.   \label{Fig1}}
\end{figure}

We suppose that two light sources A and B and a detector are fixed on a carrier.
Light source A is located at a distance $L$ on the right of the detector, whereas light source B is located on the left of the detector. 
The schematic is shown in Figure \ref{Fig1}(a). 
We know that no redshift is observed by the detector when the carrier undergoes an inertial motion. This would be different when the carrier undergoes a non-inertial motion (see Figures \ref{Fig1}(b) and (c)).

%\end{multicols}
%\begin{multicols}{2}

In this section, we firstly review the redshift in a non-relativistic approximation and in Rindler coordinates.% as well as uniformly accelerated motion in relativity.

\subsection{Non-relativistic redshift for accelerated observers}

As time is absolute in non-relativistic kinematics, the frequency of light is universal in different reference frames. This indicates that a redshift calculated in a laboratory reference frame is equal to that calculated in the reference frame for a moving detector. %In a non-relativistic approximation, we can calculate redshift in laboratory reference frame to study what the detector observes.

We consider that light source B is assigned to the left of the detector at a distance of $L$. The carrier undergoes a uniformly accelerated motion to the right. This is shown in Figure \ref{Fig1}(b). The source emits a photon at $t'$, and the detector observes the photon at $t$. In the non-relativistic approximation that $a t \ll 1 $ and $L \ll 1/a$, the processes can be formulated as
\begin{equation}
(t - t') - \dfrac{1}{2} a (t^2 - t^{\prime 2}) = L, \label{1}
\end{equation}
where $a$ is the acceleration of the carrier. There is a difference
in the time intervals of emitted and received photons. Using Eq.(\ref{1}), we have the
ratio of the time intervals,
\begin{equation}
\dfrac{\Delta t}{\Delta t'} = \dfrac{1 - a   t}{1 - a   t'} =
\dfrac{\sqrt{(1 - a   t)^2 - 2 a L}  }{1 - a   t}
. \label{2}
\end{equation}
From Eq.~(\ref{2}), the redshift, $z_-$, is given by 
\begin{equation}
z_- \equiv \dfrac{\Delta t}{\Delta t'} - 1 = \dfrac{a   L}{(1 -
	a   t)^2} +\mathcal{O} ((a L)^2) . \label{3}
\end{equation}
It shows that the redshift is time dependent and will increase with
time. Likewise, we consider light source A located on the right of the detector. It
will observe a blueshift from the source, which is given by
\begin{equation}
z_+ \approx - \dfrac{a   L}{(1 + a   t)^2} .
\end{equation}
This shows that the blueshift is also time dependent. With time, the
blueshift would become lower. The redshift and blueshift drift are illustrated in Figure \ref{Fig1}  from process (b) to (c). In non-relativistic kinematics, this should be understood as a Doppler effect, because there is a difference in the velocities % between light source and detector, 
when the ray is emitted and received. However, in the reference frames for the carrier, the difference in the velocities might not be perceived. The redshift drift  in the accelerated frames should be understood as a relativistic effect. We will show this in section \ref{IV}. The situation is  similar to the understanding of the expansion of the universe.

In most cases where $a   t \ll c = 1$, the redshift and the blueshift would reduce to the most common version, namely, $z_{\pm} = \mp \dfrac{a     L}{c^2}$.

\subsection{Redshift in Rindler coordinate}

In relativity, a uniformly accelerated motion is  described by a constant norm
of a four-acceleration for a worldline. %The motion must belong to a plane of space-time. 
In the $t$--$x$ diagram, a uniformly accelerated motion
is  a hyperbolic motion, as  the trajectory of the uniformly
accelerated motion is a hyperbola, which can be of the form,
\begin{equation}
x^2 - t^2 = \dfrac{1}{a} .
\end{equation}
The hyperbolic motion can be described by the following equations:
\begin{equation}
\left\{\begin{array}{lll}
\dfrac{ {\rm{d}} u^0}{ {\rm{d}} \tau} & = & a   u^1,\\
\dfrac{ {\rm{d}} u^1}{ {\rm{d}} \tau} & = & a   u^0,
\end{array}\right. \label{6}
\end{equation}
where $u^0 \equiv  {\rm{d}} t /  {\rm{d}} \tau$, $u^1 \equiv  {\rm{d}} x /  {\rm{d}}
\tau$, and $\tau$ is the proper time. Using the normalization condition of $u^{\mu}$,
one can find that the norm of four-acceleration $ {\rm{d}} u^{\mu} /  {\rm{d}} \tau$ is a
constant  $a$.  A solution of Eq.(\ref{6}) can be obtained,
\begin{equation}
\left\{\begin{array}{lll}
u^0 & = & \cosh (a   \tau),\\
u^1 & = & \sinh (a \tau).
\end{array}\right.  \label{7}
\end{equation}
Under specific initial condition, the parametrized trajectory of a uniformly accelerated motion can be of the form,
\begin{equation}
\left\{\begin{array}{lll}
t & = & \dfrac{1}{a} \sinh (a   \tau),\\
x & = & \dfrac{1}{a} \cosh (a \tau).
\end{array}\right.  \label{8}
\end{equation}

Another point of view for a uniformly accelerated motion in relativity is from
electrodynamics \cite{gourgoulhon_auth._special_2013}. The equations of motion for a
charged particle are of the form,
\begin{equation}
\dfrac{ {\rm{d}} u^{\mu}}{ {\rm{d}} \tau} = \dfrac{q}{m} F^{\mu}_{\hspace{0.5em}
	\nu} u^{\nu}, \label{9}
\end{equation}
where $F^{\mu}_{\hspace{0.5em} \nu}$ is the electromagnetic tensor and $m$ and $q$ are
static mass and charge of a particle, respectively. We consider a uniform electric field in the
direction of $x$-axis. For simplicity, we ignore other spatial coordinates. The
potential is given by $ A_{\mu} = (E_0 x, 0)$,
where $E_0$ is the strength of the electric field. From the potential, the
electromagnetic tensor is of the form,
\begin{equation}
F^{\mu}_{\hspace{0.5em} \nu} = \eta^{\mu \sigma} F_{\sigma \nu} =
\left(\begin{array}{cc}
& E_0\\
E_0 & 
\end{array}\right).
\end{equation}
Eq.~(\ref{9}) can be rewritten as
\begin{equation}
\left\{\begin{array}{lll}
\dfrac{ {\rm{d}} u^0}{ {\rm{d}} \tau} & = & \dfrac{E_0 q}{m} u^1,\\
\dfrac{ {\rm{d}} u^1}{ {\rm{d}} \tau} & = & \dfrac{E_0 q}{m} u^0.
\end{array}\right. 
\end{equation}
For charged particles, the equations of motion are shown to be the same as those for a
hyperbolic motion with acceleration $a = \dfrac{E_0 q}{m}$. 

A Rindler coordinate might be the most commonly used uniformly accelerated frame. %The similarity between Rindler coordinate and Kruskal coordinate of Schwarzschild space-time inspired the Hawking radiation \cite{rindler_kruskal_1966,unruh_notes_1976}.
 The metric %of Rindler coordinate 
 is given by
\begin{equation}
{\rm{d}} s^2 = - X^2  {\rm{d}} T^2 +  {\rm{d}} X^2 +  {\rm{d}} Y^2 +  {\rm{d}} Z^2 .
\end{equation}
The coordinate transformation between inertial frames and Rindler coordinates is of the form,
\begin{equation}
\left\{\begin{array}{lll}
t & = & X   \sinh (a   T),\\
x & = & X   \cosh (a   T),\\
y & = & Y,\\
z & = & Z.
\end{array}\right. 
\end{equation}
From Eq.~(\ref{8}), the coordinate transformation suggests $T \sim \tau$.
Namely, the coordinate time of a uniformly accelerated frame has a similar status of proper time for
co-moving observers. Form this point of view, there are other
coordinates that may be regarded as uniformly accelerated frames. The general transformation between 
Rindler coordinates and inertial frames is given by
\begin{equation}
\left\{\begin{array}{lll}
t & = & f (X) \sinh (a   T),\\
x & = & f (X) \cosh (a   T),\\
y & = & Y,\\
z & = & Z.
\end{array}\right.
\end{equation}
where $f (X)$ could be understood as different rulers of space. If
$f (X) = \dfrac{1}{a} + X$, they are the so-called Møller coordinates. Moreover, if $f (X) =
\dfrac{1}{a} e^{a   X}$, they are the so-called Lass coordinates \cite{lass_accelerating_1963}. In general, the
metrics are of the form,
\begin{equation}
{\rm{d}} s^2 = -a^2 f^2  {\rm{d}} T^2 + (f')^2  {\rm{d}} X^2 +  {\rm{d}} Y^2 +  {\rm{d}} Z^2
.
\end{equation}
However, it should be noted  that the norms of four-accelerations are not the same for different
co-moving observers. They depend on coordinate  $X$ of the observers. For example, if
$f = \dfrac{1}{a} + X$, the norm of the four-acceleration is given by
\begin{equation}
\sqrt{g_{\mu \nu} a^{\mu} a^{\nu}} = \dfrac{1}{f (X)} = \dfrac{a}{1 + a
	X} . \label{16}
\end{equation}
If we wish to construct a uniformly accelerated frame %with the congruence 
based on a picture of a charged particle in a uniform electric field, the norms of four-accelerations should be constant for all the observers located at arbitrary positions. This led Huang and Guo \cite{huang_new_2006} to construct new types of uniformly
accelerated frames, in these new frames, the norms of accelerations for co-moving observers are
the same constant, $a$. Another understanding of Eq.~(\ref{16}) was given by
MTW \cite{misner_gravitation_1973}, who derived Møller coordinates with the hypothesis of locality.  At the location where $X \ll a^{- 1}$, the norms of four-accelerations for different co-moving observers are nearly the same constants. This indicates that Rindler coordinates are in fact local frames.

Observables in Rindler coordinates are redshifts between co-moving observers
\cite{landsberg_gravitational_1976}, which can be given by
\begin{equation}
z_{\pm}= \sqrt{\dfrac{g_{T   T} (X)}{g_{T   T} (X')}} =
\dfrac{1 + a   X}{1 + a   X'} - 1 \approx a (X - X'), \label{17}
\end{equation}
where $X$ and $X'$ are the fixed positions of the detector and the source, respectively. The redshift
is time-independent%. It seems that the redshift in Rindler coordinate 
and is different from that calculated in the non-relativistic approximation, Eq.~(\ref{3}).

Huang \cite{huang_uniformly_2008} suggested that this difference is originated from Eq.
(\ref{16}) that norms of 4-accelerations are not the same constant for all co-moving
observers. For their new kind of uniformly accelerated reference frames with
the same accelerations for co-moving observers \cite{huang_new_2006}, the redshift was
shown to be time-dependent.

\section{Uniformly Accelerated Reference Frame}\label{III}

The construction of the alternative uniformly accelerated references was based on 3+1 formalism in adapted coordinates, 
which is also different from the uniformly accelerated reference frames suggested by Huang \cite{huang_new_2006}.

For simplicity, the accelerations of the frames are set along the $x$ direction.
A coordinate transformation between the uniformly accelerated frames and the inertial frames is expected in the form,
\begin{equation}
\left\{\begin{array}{lll}
T & = & T (t, x),\\
X & = & X (t, x),\\
Y & = & y,\\
Z & = & z,
\end{array}\right. \label{18}
\end{equation}
where $T, X, Y$ and $Z$ are the coordinates of  the uniformly accelerated frames, and $t,
x, y$, and $z$ are the coordinates of the inertial frames. %The transformation indicates that differential form $ {\rm{d}} X^{\mu}$ are integrable and $ {\rm{d}}^2 X^{\mu} = 0$. 
For an accelerated frame, we expect
that a type of principle of relativity should be satisfied.
\begin{description}
	\item[Axiom 1] The co-moving observers in the uniformly accelerated frames
	undergo uniformly accelerated motions with respect to the inertial frames.
	
	\item[Axiom 2] The co-moving observers in the inertial frames undergo uniformly
	accelerated motions with respect to the uniformly accelerated frames.
\end{description}
A uniformly accelerated motion in the axioms is formulated as Eq.~(\ref{6}). We would use axiom 1 for constructing the uniformly accelerated reference
frames. This suggests means that the uniformly accelerated observers defined in the inertial frames should move attached to the accelerated reference frames. In the
following, we verify axiom 2 by the fact that the geodesics in uniformly accelerated
frames can be formulated as uniformly accelerated motions.

\subsection{Construction of uniformly accelerated reference frame}

In the 3+1 formalism, four-velocities $u$ of accelerated observers are normal
vectors of a space-like hypersurface $\Sigma_T$, which is formulated as
\begin{equation}
u_{\mu}  {\rm{d}} x^{\mu} = - N  {\rm{d}} T,
\end{equation}
where $N$ is the so-called lapse function and $N > 0$. A uniformly accelerated frame is adapted to $u$. Moreover we set that the $u$ is along the direction of the $x$-axis, namely, $u^2 = u^3 = 0$ for simplicity. $T$ is the coordinate time of the
uniformly accelerated frames. The transformation for $ {\rm{d}} T$ can be written as
\begin{equation}
{\rm{d}} T = \dfrac{u^0}{N}  {\rm{d}} t - \dfrac{u^1}{N}  {\rm{d}} x. \label{20}
\end{equation}
This suggests that $\partial_0 T = \dfrac{u^0}{N}$ and $\partial_1 T =
\dfrac{u^1}{N}$. %One can also understand the $N$ via,
%\begin{equation}
%  \dfrac{ {\rm{d}} T}{ {\rm{d}} \tau} = u^0 \partial_0 T + u^1 \partial_1 T =
%  \dfrac{1}{N},
%\end{equation}
%where $u^T \equiv \dfrac{ {\rm{d}} T}{ {\rm{d}} \tau}$, which is time component of
%the accelerated observers with respect to uniformly accelerated frame.
$N$ as an integrating factor ensuring that differential form $ {\rm{d}} T$ is
integrable. Because $ {\rm{d}}^2 = 0$, Eq.~(\ref{20}) leads to
\begin{equation}
\partial_1 \left( \dfrac{u^0}{N} \right) + \partial_0 \left( \dfrac{u^1}{N}
\right) = 0. \label{22}
\end{equation}
The transformation for $X$ at present is arbitrary, which can be written as
\begin{equation}
{\rm{d}} X = \partial_0 X  {\rm{d}} t + \partial_1 X  {\rm{d}} x. \label{23}
\end{equation}
For coordinates $Y$ and $Z$, the transformation are respectively given by ${\rm{d}} Y  =   {\rm{d}} y$ and ${\rm{d}} Z  =   {\rm{d}} z.$
%\begin{equation}
%\left\{\begin{array}{lll}
%{\rm{d}} Y & = &  {\rm{d}} y,\\
%{\rm{d}} Z & = &  {\rm{d}} z.
%\end{array}\right. 
%\end{equation}
From Eqs.~(\ref{20}) and (\ref{23}), we obtain the inverse transformations for $ {\rm{d}} t$ and
$ {\rm{d}} x$,
\begin{equation}
\left\{\begin{array}{lll}
{\rm{d}} t & = & \dfrac{N}{u^0 \partial_1 X + u^1 \partial_0 X} \left(
\partial_1 \text{} X  {\rm{d}} T + \dfrac{u^1}{N}  {\rm{d}} X \right),\\
{\rm{d}} x & = & \dfrac{N}{u^0 \partial_1 X + u^1 \partial_0 X} \left( -
\partial_0 X  {\rm{d}} T + \dfrac{u^0}{N}  {\rm{d}} X \right).
\end{array}\right.  \label{25}
\end{equation}
With the transformations, one can obtain the metric of the uniformly accelerated
reference frames,
%\end{multicols}
\begin{eqnarray}
{\rm{d}} s^2 & = & -  {\rm{d}} t^2 +  {\rm{d}} x^2 +  {\rm{d}} y^2 +  {\rm{d}} z^2
\nonumber\\
%& = & \dfrac{N^2}{(u^0 \partial_1 X + u^1 \partial_0 X)^2} \left( - \left(
%\partial_1 X  {\rm{d}} T + \dfrac{u^1}{N}  {\rm{d}} X \right)^2 + \left( -
%\partial_0 X  {\rm{d}} T + \dfrac{u^0}{N}  {\rm{d}} X \right)^2 \right) +  {\rm{d}}
%Y^2 +  {\rm{d}} Z^2 \nonumber\\
& = & \dfrac{N^2}{(u^0 \partial_1 X + u^1 \partial_0 X)^2} \left( (-
(\partial_1 X)^2 + (\partial_0 X)^2)  {\rm{d}} T^2 \right. \nonumber \\
 & & \left. + \dfrac{1}{N^2}  {\rm{d}} X^2 
-   \dfrac{2}{N} (u^0 \partial_0 X  + u^1 \partial_1 X)  {\rm{d}} T  {\rm{d}} X \right)  \nonumber  \\
& & +  {\rm{d}} Y^2 +  {\rm{d}} Z^2 .  \label{26}
\end{eqnarray}
%\begin{multicols}{2}
From axiom 1, accelerated observer $u$ should be a co-moving observer of the uniformly accelerated frames, namely,
\begin{equation}
u^0 \partial_0 X + u^1 \partial_1 X = \dfrac{ {\rm{d}} X}{ {\rm{d}} \tau} = 0.
\label{27}
\end{equation}
Moreover, if set $\gamma = \dfrac{1}{u^0 \partial_1 X + u^1 \partial_0 X}$,
by making use of Eq.~(\ref{27}), we can rewrite $\partial_0 X$ and $\partial_1 X$
as
\begin{equation}
\left\{\begin{array}{lll}
\partial_0 X & = & - \dfrac{u^1}{\gamma},\\
\partial_1 X & = & \dfrac{u^0}{\gamma}.
\end{array}\right. \label{28}
\end{equation}
Using Eqs.~(\ref{25}), (\ref{26}), and (\ref{28}), we know that the metric is reduced to
\begin{equation}
{\rm{d}} s^2 = - N^2  {\rm{d}} T^2 + \gamma^2  {\rm{d}} X^2 +  {\rm{d}} Y^2 +  {\rm{d}}
Z^2 . \label{29}
\end{equation}
In the adapted coordinates, condition axiom 1 guarantees that the metric is always diagonal. This is different from the uniformly accelerated frames suggested by Huang \cite{huang_new_2006}.

One may wonder how much this is related to the 3+1 formalism.
In general, a metric in the 3+1 formalism can be written as
\begin{equation}
{\rm{d}} s^2 = - N^2  {\rm{d}} T^2 + \gamma_{i   j} (\beta^i  {\rm{d}} T +
{\rm{d}} X^i) (\beta^j  {\rm{d}} T +  {\rm{d}} X^j), \label{30}
\end{equation}
where $\gamma_{i   j}$ is the reduced metric and $\beta^i$ is the so-called
shift function. Owing to axiom 1, $\beta^i$ is shown to vanish. As we know $\beta^X = N   u^X = N \dfrac{ {\rm{d}} X}{ {\rm{d}} \tau} = 0$
%\begin{equation}
%\beta^X = N   u^X = N \dfrac{ {\rm{d}} X}{ {\rm{d}} \tau} = 0.
%\end{equation}
and $u^Y = u^Z = 0$, this leads to $\beta^i = 0$. We can rewrite the metric in 
Eq.~(\ref{30}) as
\begin{equation}
{\rm{d}} s^2 = - N^2  {\rm{d}} T^2 + \gamma_{i   j}  {\rm{d}} X^i  {\rm{d}} X^j \label{32}
.
\end{equation}
%The metric is diagonal. 
Comparing it with Eq.~(\ref{29}),
one can find that the reduced metric, $\gamma_{i   j}$, in the
accelerated frames is of the form,
\begin{equation}
\gamma_{i   j} = \left(\begin{array}{ccc}
\gamma^2 &  & \\
& 1 & \\
&  & 1
\end{array}\right) .
\end{equation}
As $\gamma$ also functions as an integrating factor, we prefer $\gamma$ to
$\gamma_{i   j}$ in our derivation.

In the metric in Eq.~(\ref{29}), there are two unknown fields, $g_{00}$ and
$g_{11}$, which depend on the choice of the four-velocity, $u$.Here, we consider 
uniformly accelerated reference frames. Namely, $u$ %describes co-moving observers that undergo a uniformly accelerated motion. From the second equation of Eqs.~(\ref{6}), the accelerated motion 
is decribed by Eq.~(\ref{6}), which can be rewritten as
\begin{equation}
\dfrac{ {\rm{d}} u^1}{ {\rm{d}} \tau} %= \dfrac{ {\rm{d}} T}{ {\rm{d}} \tau} \partial_T u^1
= \dfrac{1}{N} \partial_T u^1 = a   u^0 . \label{34}
\end{equation}
By making use of Eq.~(\ref{28}), we rewrite Eq.~(\ref{22}) in terms of the coordinates, $(T,X,Y,Z)$,
%of uniformly accelerated frame as
\begin{equation}
\partial_1 \left( \dfrac{u^0}{N} \right) + \partial_0 \left( \dfrac{u^1}{N}
\right) = \dfrac{1}{\gamma} \partial_X \dfrac{1}{N} + \dfrac{a}{N} = 0.
\label{35}
\end{equation}
As $ {\rm{d}}^2 X = 0$, it leads to an equation as follows:
\begin{equation}
\partial_1 \left( \dfrac{u^1}{\gamma} \right) + \partial_0 \left(
\dfrac{u^0}{\gamma} \right) = \dfrac{1}{N} \partial_T \left( \dfrac{1}{\gamma}
\right) + \dfrac{1}{u_0 \gamma^2} \partial_X u^1 = 0. \label{36}
\end{equation}
We rearrange Eqs.~(\ref{34}), (\ref{35}), and (\ref{36}) as
\begin{equation}
\left\{\begin{array}{lll}
\partial_X N & = & a   N \gamma,\\
\partial_T G & = & a   N,\\
\partial_X G & = & \dfrac{1}{N} \partial_T \gamma,
\end{array}\right. \label{37}
\end{equation}
where we set $\partial G \equiv \dfrac{1}{u_0} \partial u^1$. As $u^{\mu}
u_{\mu} = - 1$, it leads to $G = {\rm{arcsinh}} (u^1)$.
%\begin{equation}
%G = {\rm{arcsinh}} (u^1) . \label{38}
%\end{equation}
%At the same time, we notice solution of hyperbolic motion, the second equation of Eqs.~(\ref{7}), which suggests that
Associating it with Eq.~(\ref{7}), we find
\begin{equation}
G = a \tau . \label{39}
\end{equation}
Then, the Eq.~(\ref{37}) can be rewritten in a natural manner,
\begin{equation}
\left\{\begin{array}{lll}
\partial_X N & = & a   N \gamma,\\
\partial_T \tau & = & N,\\
\partial_X \tau & = & \dfrac{1}{a   N} \partial_T \gamma.
\end{array}\right.  \label{40}
\end{equation}
The solutions of Eq.~(\ref{40}) provide an explicit metric of the uniformly
accelerated reference frames. The expression of the coordinate transformation depends on $N$, $\gamma$, and proper time $\tau$. From Eqs.~(\ref{20}), (\ref{28}), and (\ref{39}), the coordinate transformation,
which takes the form of Eq.~(\ref{18}), can be derived from
\begin{equation}
\left\{\begin{array}{lll}
{\rm{d}} T & = & \dfrac{\cosh   (a \tau)}{N}  {\rm{d}} t - \dfrac{\sinh
	(a \tau)}{N}  {\rm{d}} x,\\
{\rm{d}} X & = & - \dfrac{\sinh   (a \tau)}{\gamma}  {\rm{d}} t +
\dfrac{\cosh (a \tau)}{\gamma}  {\rm{d}} x, \\
{\rm{d}} Y & = &   {\rm{d}} y, \\
{\rm{d}} Z & = &   {\rm{d}} z.
\end{array}\right. 
\end{equation}

Besides the diagonal, there are other features of the metric from Eq.~(\ref{40}). Firstly, the metric must depend on
coordinate time $T$. %If inserting $\gamma = \gamma (X)$, one might find the constraint equations would be contradictory. 
This indicates that the Rindler metric can not be included in
our uniformly accelerated frames. Secondly, $N = 1$ is not permitted. This means that coordinate time $T$ of the uniformly accelerated
frames is not the proper time for co-moving observers. This could be understood by analogy. In the Schwarzschild
space-time, one might not require that the coordinate time for a co-moving observer be a
proper time, because there is gravity. In the uniformly accelerated frames, also this is the case, because there is a fiction force.

From Eq.~(\ref{40}), we may prove that the metric (Eq.~(\ref{29})) is a solution of
vacuum Einstein equations. We can start to check this by calculations of the
connection,
\begin{eqnarray}
\left\{\begin{array}{lll}
\Gamma^T_{T   T} & = & \dfrac{\partial_T N}{N}, \\
\Gamma^T_{T   X} & = & \dfrac{\partial_X N}{N} = a   \gamma, \\
\Gamma^T_{X   X} & = & \dfrac{\gamma \partial_T \gamma}{N^2}, \\
\Gamma^X_{T   T} & = & \dfrac{N \partial_X N}{\gamma^2} = \dfrac{a
	N^2}{\gamma}, \\
\Gamma^X_{T   X} & = & \dfrac{\partial_T \gamma}{\gamma}, \\
\Gamma^X_{X   X} & = & \dfrac{\partial_X \gamma}{\gamma}, \\
\rm{others} & = & 0. 
\end{array}\right. 
\end{eqnarray}
Non-trivial components of the Ricci tensor are given by
\begin{eqnarray}
\left\{\begin{array}{l}
R_{T   T}  =  \dfrac{1}{\gamma^3} \left( N   \gamma^2 \partial_X \left(
\dfrac{\partial_X N}{\gamma} \right) - N   \gamma^2 \partial_T \left(
\dfrac{\partial_T \gamma}{N} \right) \right),  \\
R_{X   X} %& = & \dfrac{1}{N^3} \left( \dfrac{N^2 \partial_X \gamma
%  \partial_X N}{\gamma} - N^2 \partial_X^2 N + \gamma (- \partial_T \gamma
%  \partial_T N + N \partial_T^2 \gamma) \right) \\
 =  \dfrac{1}{N^3} \left( - \gamma N^2 \partial_X \left( \dfrac{\partial_X
	N}{\gamma} \right) + \gamma N^2 \partial_T \left( \dfrac{\partial_T
	\gamma}{N} \right) \right) . 
\end{array}\right. 
\end{eqnarray}
From Eq.~(\ref{40}), one has
\begin{eqnarray}
\left\{\begin{array}{lll}
\partial_T \left( \dfrac{\partial_T \gamma}{N} \right) & = & a^2 N    \gamma, \\
\partial_X \left( \dfrac{\partial_X N}{\gamma} \right) & = & a^2 N \gamma . 
\end{array}\right. 
\end{eqnarray}
This causes the Ricci tensor to be zero,
\begin{equation}
R_{\mu \nu} = 0.
\end{equation}
Namely, the metric of the uniformly accelerated frames is a solution of  vacuum Einstein equations. The checking process seems trivial, as the Einstein equation always allows coordinate transformation as a gauge symmetry.

\subsection{Inertial motion in uniformly accelerated frame}

In subsection A, we have constructed the metric of the uniformly accelerated frames with
axiom 1. The axiom presents an equivalent description for uniformly accelerated
motions in different reference frames. On the other side, an inertial motion also requires an equivalent description. This suggests that the inertial
motion should be formulated as a uniformly accelerated motion in the view of a uniformly
accelerated frame.

We consider the inertial motion in the uniformly accelerated frame at the $T
$--$X$ plane,
\begin{equation}
\left(\begin{array}{c}
\dfrac{{\rm{d}} T}{{\rm{d}} \tau_0}\\
\dfrac{{\rm{d}} X}{{\rm{d}} \tau_0}
\end{array}\right)  =  \left(\begin{array}{cc}
\dfrac{\cosh  (a \tau)}{N} & - \dfrac{\sinh  (a \tau)}{N}\\
- \dfrac{\sinh  (a \tau)}{\gamma} & \dfrac{\cosh  (a
	\tau)}{\gamma}
\end{array}\right) \left(\begin{array}{c}
\dfrac{{\rm{d}} t}{{\rm{d}} \tau_0}\\
\dfrac{{\rm{d}} x}{{\rm{d}} \tau_0}
\end{array}\right)
\label{53},
\end{equation}
where $(\dfrac{{\rm{d}} t}{{\rm{d}} \tau_0},\dfrac{{\rm{d}} x}{{\rm{d}} \tau_0})$ is a constant velocity vector, $\tau_0$ is the proper time of co-moving observers in the inertial frame, which
is distinguished from $\tau$ proper time for the uniformly accelerated observers.
From Eqs.~(\ref{40}) and (\ref{53}), one can obtain
\begin{equation}
\left\{\begin{array}{lll}
\dfrac{ {\rm{d}}}{ {\rm{d}} \tau} \left( N \dfrac{ {\rm{d}} T}{ {\rm{d}} \tau_0}
\right) = - a   \gamma \dfrac{ {\rm{d}} X}{ {\rm{d}} \tau_0} , \\
\dfrac{ {\rm{d}}}{ {\rm{d}} \tau} \left( \gamma \dfrac{ {\rm{d}} X}{ {\rm{d}} \tau_0}
\right) = - a   N \dfrac{ {\rm{d}} T}{ {\rm{d}} \tau_0} . \label{57}
\end{array}\right. 
\end{equation}
If we set $  \upsilon^T =  N \dfrac{ {\rm{d}} T}{ {\rm{d}} \tau_0}$ and $ \upsilon^X  =  \gamma \dfrac{ {\rm{d}} X}{ {\rm{d}} \tau_0}$,
%\begin{eqnarray}
% \left\{\begin{array}{lll}
% \upsilon^T & = & N \dfrac{ {\rm{d}} T}{ {\rm{d}} \tau_0}  \label{58}\\
% \upsilon^X & = & \gamma \dfrac{ {\rm{d}} X}{ {\rm{d}} \tau_0},  \label{59}
%   \end{array}\right. 
%\end{eqnarray}
then Eq.~(\ref{57}) reduces to
\begin{equation}
\left\{\begin{array}{lll}
\dfrac{ {\rm{d}} \upsilon^T}{ {\rm{d}} \tau} & = & - a   \upsilon^X,\\
\dfrac{ {\rm{d}} \upsilon^X}{ {\rm{d}} \tau} & = & - a \upsilon^T.
\end{array}\right.  \label{60}
\end{equation}
From Eq.~(\ref{60}), the inertial motion in the view of the uniformly accelerated
frames can be formulated as a hyperbolic motion with a reverse acceleration. Using Eq.~(\ref{53}), one can verify  that $\dfrac{{\rm{d}} T}{{\rm{d}} \tau_0}$ and $\dfrac{{\rm{d}} X}{{\rm{d}} \tau_0}$ satisfy geodesic equations in the accelerated frames. All these
indicate that axiom 2 is verified. In addition, the redefinitions of the
covariant velocities $(\upsilon^T, \upsilon^X)$ are insightful. In
curvilinear coordinates, this is exactly the standard definition of a vector,
where $N, \gamma$ are the so-called Lamé coefficients. Moreover, in the tetrad
formalism, one may find $\upsilon^a = e_{\mu}^{\hspace{0.5em} a} u^{\mu}$.

\subsection{Features of uniformly accelerated reference frames}

\subsubsection{Frenet--Serret frames}

The Frenet--Serret frames describe the evolution of the frames along
worldline of the observers. It can be generally written as
\begin{equation}
\dfrac{D  }{ {\rm{d}} \tau} \left(\begin{array}{c}
e_0\\
e_1\\
e_2\\
e_3
\end{array}\right) = \left(\begin{array}{cccc}
& \kappa &  & \\
\kappa &  & \tau & \\
& - \tau &  & b\\
&  & - b & 
\end{array}\right) \left(\begin{array}{c}
e_0\\
e_1\\
e_2\\
e_3
\end{array}\right),
\end{equation}
where $\dfrac{D}{{\rm{d}}\tau}$ is the covariant derivative, $e_{\mu}$ represent the vector bases of the frames, the $\kappa$, $\tau$, and $b$
are the so-called curvature and torsions of worldline of the observers in the Lorentz manifold, respectively.

Our uniformly accelerated frames can be expressed in terms of the Frenet--Serret
frames, in the tetrad formalism. Firstly, we rewrite coordinate bases $\partial_{\mu}$
as tetrads $e_a $, which are formulated as $  e_a = e_a^{\hspace{0.5em} \mu} \partial_{\mu} $.
%\begin{equation}
% e_a = e_a^{\hspace{0.5em} \mu} \partial_{\mu} .
%\end{equation}
In the uniformly accelerated frames, the tetrads can be given by
\begin{equation}
\left\{\begin{array}{lll}
e_0 & = & \dfrac{1}{N} \partial_T,\\
e_1 & = & \dfrac{1}{\gamma} \partial_X,\\
e_2 & = & \partial_Y,\\
e_3 & = & \partial_Z.
\end{array}\right. 
\end{equation}
%The quantities $\kappa, \tau$ and $b$ are obtained via the following relation,
%\begin{equation}
%  \dfrac{D   e_0}{ {\rm{d}} \tau} = \dfrac{D   u}{ {\rm{d}} \tau} = a
%    e_1,
%\end{equation}
%\begin{equation}
%  \dfrac{D   e_1}{ {\rm{d}} \tau} = \left(\begin{array}{c}
%    \dfrac{1}{N} \left( \partial_T 0 + \Gamma^T_{T   X} \dfrac{1}{\gamma}
%    \right)\\
%    \dfrac{1}{N} \left( \partial_T \dfrac{1}{\gamma} + \Gamma^X_{T   X}
%    \dfrac{1}{\gamma} \right)\\
%    0\\
%    0
%  \end{array}\right) = a   e_0,
%\end{equation}
%And $\dfrac{D   e_2}{ {\rm{d}} \tau} = \dfrac{D   e_3}{ {\rm{d}} \tau} =0$. 
One can find that $\kappa = a$, $\tau = b = 0$ for our uniformly
accelerated frames, namely,
\begin{equation}
\dfrac{D  }{ {\rm{d}} \tau} \left(\begin{array}{c}
e_0\\
e_1\\
e_2\\
e_3
\end{array}\right) = \left(\begin{array}{cccc}
0 & a &  & \\
a & 0 &  & \\
&  & 0 & \\
&  &  & 0
\end{array}\right) \left(\begin{array}{c}
e_0\\
e_1\\
e_2\\
e_3
\end{array}\right).
\end{equation}
The curvature of the moving frames is exactly the constant acceleration, $a$, in
our uniformly accelerated frames, whereas the Rindler coordinates can not be
described in terms of Frenet--Serret frames with tetrads.

\subsubsection{Kinematical quantities}

The congruence of co-moving observers $u$ indicates a deformation of the space--time. The difference
between the worldlines of co-moving observers can be described in terms of deviation vector $\chi^{\mu}$, which satisfies
\begin{equation}
[u, \chi]^{\mu} = u^{\nu} \nabla_{\nu} \chi^{\mu} - \chi^{\nu} \nabla_{\nu}
\chi^{\mu} = 0~. \label{67}
\end{equation}
For the spatial distance of $\chi^{\mu}$, namely, $\tilde{\chi}^{\mu} =
\gamma^{\mu}_{\nu} \chi^{\nu}$, the evolution of $\tilde{\chi}^{\mu}$
indicates the spatial deformation of the reference frames,
\begin{equation}
\dfrac{\tilde{D} \tilde{\chi}^{\mu}}{ {\rm{d}} \tau} = \chi^{\mu} \left(
\dfrac{1}{2} \theta \gamma_{\nu}^{\mu} + \sigma^{\mu}_{\nu} +
w^{\mu}_{\hspace{0.5em} \nu} \right), \label{68}
\end{equation}
where $\dfrac{\tilde{D}}{ {\rm{d}} \tau} = \gamma^{\mathord{*}} \nabla_u$ is
the spatial covariant derivation with respect to $u$. % along a co-moving observer derived from 3+1 foliation. 
$\theta, \sigma^{\mu}_{  \nu}$, and
$w^{\mu}_{\hspace{0.5em} \nu}$ are the so-called kinematical quantities and named
after the expansion scalar, shear tensor, and rotation tensor, respectively. 

In the accelerated frames, co-moving observers $u$ are given by
\begin{equation}
u_{\mu}  {\rm{d}} X^{\mu} = (- N, 0, 0, 0)~.
\end{equation}
The covariant derivative of observers $u$ can be decomposed into acceleration, expansion
scalar, shear, and rotation tensor, 
\begin{eqnarray}
\nabla_{\nu} u_{\mu} %& = & (- u_{\nu} u^{\sigma} + \gamma^{\sigma}_{\nu}) (-
% u_{\mu} u^{\rho} + \gamma^{\rho}_{\mu}) \nabla_{\sigma} u_{\rho} \nonumber\\
& = & - u_{\nu} a_{\mu} + \dfrac{1}{3} \theta \gamma_{\mu \nu} + \sigma_{\mu
	\nu} + w_{\mu \nu},  \label{70}
\end{eqnarray}
where,
\begin{eqnarray}
a^{\mu} & = & \nabla_u u^{\mu} \label{71},
\end{eqnarray}
\begin{eqnarray}
\theta & = & \nabla_{\mu} u^{\mu} ,
\end{eqnarray}
\begin{eqnarray}
\sigma_{\mu \nu} & = & \dfrac{1}{2} (\nabla_{\mu} u_{\nu} + \nabla_{\nu}
u_{\mu} + u_{\mu} \nabla_u u_{\nu} + u_{\nu} \nabla_u u_{\mu}) - \dfrac{1}{3}
\theta \gamma_{\mu \nu} ,
\end{eqnarray}
\begin{eqnarray}
w_{\mu \nu} & = & \dfrac{1}{2} \gamma^{\sigma}_{\mu} \gamma^{\rho}_{\nu}
(\nabla_{\sigma} u_{\rho} - \nabla_{\rho} u_{\sigma}) . 
\end{eqnarray}
From Eqs.~(\ref{32}) and (\ref{71}), the accelerations of co-moving observers are given by $a^{\mu} = \delta^{\mu}_1 \dfrac{a}{\gamma}$. 
%\begin{equation}
%a^{\mu} = \delta^{\mu}_1 \dfrac{a}{\gamma}~,
%\end{equation}
The norms of the accelerations are 
\begin{equation}
| a | \equiv \sqrt{g_{\mu \nu} a^{\mu} a^{\nu}} = a~.
\end{equation}
This shows that the accelerations of any co-moving observers in our uniformly
accelerated frames are the same constant acceleration, $a$, which is different from that in Rindler coordinates (Eq.~(\ref{16})).

We present the expansion scalar, shear tensor, and rotation tensor in the following:
\begin{eqnarray}
\left\{\begin{array}{lll}
w_{\mu \nu} & = & 0,  \label{79} \\
\theta & = & \dfrac{\partial_T \gamma}{N \gamma}  \label{77},\\
\sigma_{\mu \nu} & = & \dfrac{1}{3} \theta \left( \begin{array}{cccc}
0 &  &  & \\
& 2 &  & \\
&  & - 1 & \\
&  &  & - 1
\end{array} \right).  \label{78}\\
\end{array}\right. 
\end{eqnarray}
The vanished rotation tensor suggests that co-moving observers $u$ are
Eulerian observers. There is a simultaneous hypersurface $\Sigma_T$ 
orthogonal to all the co-moving observers.

From Eqs.~(\ref{67}), (\ref{68}), and (\ref{79}), the
evolution of the spatial deviation vector is given by
\begin{equation}
\dfrac{\tilde{D} \tilde{\chi}}{ {\rm{d}} \tau} = \left(\begin{array}{cccc}
0 &  &  & \\
& \theta &  & \\
&  & 0 & \\
&  &  & 0
\end{array}\right) \tilde{\chi}~.
\end{equation}
It shows that there is a spatial deformation between the co-moving worldlines in the direction of the $X$-axis. % while in spatial direction referred to $Y, Z$, there is not a deformation. 
In the uniformly accelerated frames, it can be understood as a
non-inertial effect. The fiction force can affect the deformation of the space.

Evolutions of these kinematical quantities are described by the Raychaudhuri
equations. In our uniformly accelerated frames, the equations can be deduced from  Eq.~(\ref{40}). It did not lead to any constraints for
constructing the uniformly accelerated frames.

\section{Explicit solutions}\label{IV}

Now, we try  to obtain the solutions of Eq.~(\ref{40}). Moreover, using these solutions, we calculate the redshift drift
between co-moving objects and the possible Unruh effect in the
accelerated frames. As there is nothing special in the directions
of $Y$ and $Z$-axis, we consider two-dimensional metrics for simplicity.

\subsection{Hyperbolic metric and redshift drift}

\begin{figure}[h]
\includegraphics[width=0.6\linewidth]{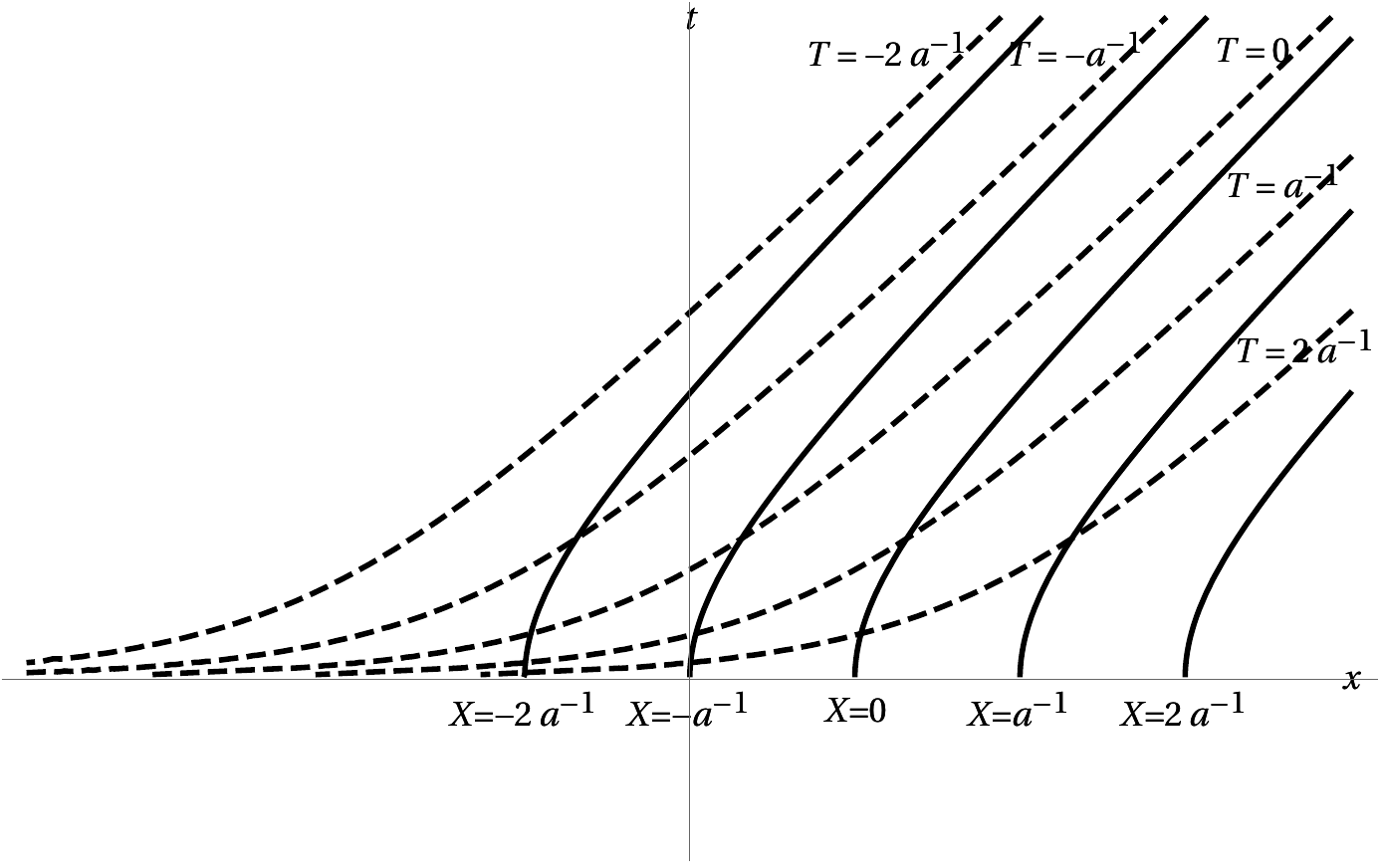}
	\caption{Coordinate lines of uniformly accelerated frames in  $t$--$x$ diagram. In case of $a>0$, accessible region for accelerated frames is that with $t>0$. \label{Fig2}}
\end{figure}

The components of the metric turn to be hyperbolic triangle functions, when the uniformly accelerated observer, $u$, is just a function of
coordinate time $t$, namely, $u = u (t)$. Associating it with Eq.~(\ref{40}),
we get the metric as
\begin{eqnarray}
{\rm{d}} s^2  =  - \dfrac{ {\rm{d}} T^2}{\sinh^2 (- a (T + X))} + \dfrac{ {\rm{d}}
	X^2}{\tanh^2 (- a (T + X))} .  \label{81}
\end{eqnarray}
As $N > 0$, it leads to $- a (T + X) > 0$. Transformation from the inertial frames
to the accelerated frames is of the form,
\begin{equation}
\left\{\begin{array}{lll}
t & = & \dfrac{1}{a   \sinh (- a (X + T))},\\
x & = & X + \dfrac{1}{a   \tanh (- a (X + T))}.
\end{array}\right.  \label{82}
\end{equation}
From Eq.~(\ref{40}), we can obtain proper time $\tau$ for the co-moving observers in the
accelerated frames, %And we can solve it as function of space-time coordinate,
\begin{equation}
\tau = \dfrac{1}{a} {\rm{arcsinh}} \left( \dfrac{1}{\sinh (- a (T + X))}
\right) = \dfrac{1}{a} {\rm{arcsinh}} (a   t) . \label{83}
\end{equation}
Eqs.~(\ref{81}) and (\ref{83}) lead to $\sinh (a \tau) = a   t
= N > 0$. If $a > 0$, accessible region of space-time is that with $\tau, t
\geqslant 0$. Namely, the reference frames undergo uniformly accelerated
motions from $t = 0$. The coordinate lines of the uniformly accelerated frames in the
$t$--$x$ plane are presented in Figure \ref{Fig2}.

%Using Eq.~(\ref{83}), we obtain velocity of a uniformly accelerated particle in
%inertial frame, which can be of the form,
%\begin{equation}
% \left\{\begin{array}{lllll}
%  u^0 & = & \sqrt{1 + (a   t)^2} & = & \cosh (a \tau)\\
%  u^1 & = & a   t & = & \sinh (a \tau)
%\end{array}\right. .
%\end{equation}
%It's exactly hyperbolic motion. Using coordinate transformation Eq.
%(\ref{82}), the particle is static with respect to the uniformly accelerated
%frame, which is given by,
%\begin{equation}
%  \left\{\begin{array}{lll}
%    u^T & = & \sinh (- a (T + X)) = \dfrac{1}{\sinh (a \tau)}\\
%    u^X & = & 0
%  \end{array}\right. .
%\end{equation}

%For a static particle in inertial frame, namely, $\dfrac{ {\rm{d}} t}{ {\rm{d}}
%\tau_0} = 1, \dfrac{ {\rm{d}} x}{ {\rm{d}} \tau_0} = 0$, one can calculate its
%velocity in uniformly accelerated frame with definition of Eq (\ref{58}) and
%(\ref{59}) as,
%\begin{eqnarray}
%  \upsilon^T & = & N \dfrac{ {\rm{d}} T}{ {\rm{d}} \tau_0} = \dfrac{1}{\tanh (- a (T +
%  X))} \nonumber\\
%  & = & \cosh (a \tau), 
%\end{eqnarray}
%\begin{eqnarray}
%  \upsilon^X & = & \gamma \tfrac{ {\rm{d}} X}{ {\rm{d}} \tau_0} = - \dfrac{1}{\sinh
%  (- a (T + X))} \nonumber\\
%  & = & - \sinh (a \tau), 
%\end{eqnarray}
%It shows that the inertial motion is hyperbolic motion in the view of
%uniformly accelerated frame.

%We assign a detector co-moving with the accelerated frame and a light source behind and co-moving with the detector. 
The metric in Eq.~(\ref{81}) can describe the reference frames of the carrier in Figure \ref{Fig1}, so that we can reconsider the redshift drift in the uniformly accelerated reference frames. The reference frames move with constant acceleration $a$ to the right, and light source B on the left of the detector co-moves with the carrier.
The source emits light that is along a null curve. By utilizing the metric, in Eq.~(\ref{81}),  we obtain the equation of trajectories of the light,
\begin{equation}
\dfrac{1}{\sinh (- a (T + X))} \pm \dfrac{1}{\tanh (- a (T + X))} \dfrac{ {\rm{d}}
	X}{ {\rm{d}} T} = 0~. \label{88}
\end{equation}
In this case, only forward-propagating light
reaches the detector. Namely, the $'' -''$ in  Eq.~(\ref{88}) is required to be
chosen. Then, the trajectories are obtained
\begin{equation}
\tanh \left( - \dfrac{a}{2} (T + X) \right) = - a (X - X') + \tanh \left( -  \dfrac{a}{2} (T' + X') \right)~, \label{89}
\end{equation}
where the detector and the source are fixed at spatial coordinates $X$ and
$X'$, where $X > X'$, respectively. From Eq.~(\ref{89}), it takes different time intervals, when two light signals are emitted and received. The ratio can be given by
\begin{equation}
\dfrac{\Delta T}{\Delta T'} = \dfrac{\cosh^2 \left( - \dfrac{a}{2} (T + X)
	\right)}{\cosh^2 \left( - \dfrac{a}{2} (T' + X') \right)}, \label{90}
\end{equation}
where $\Delta T'$ and $\Delta T$ are the emitted and received time intervals,
respectively. Using Eqs.~(\ref{81}), (\ref{83}), (\ref{89}), and (\ref{90}), we
can derive the redshift \cite{wald_general_1984} observed by the detector in the uniformly accelerated reference frames, 
\begin{eqnarray}
z_- & \equiv & \dfrac{\Delta \tau}{\Delta \tau'} - 1 = \dfrac{\sqrt{g_{T
			T} (T, X)} \Delta T}{\sqrt{g_{T   T} (T', X')} \Delta T'} -
1 \nonumber\\
& = & \dfrac{\tanh \left( - \dfrac{a}{2} (T' + X') \right)}{\tanh \left( -
	\dfrac{a}{2} (T + X) \right)} - 1 \nonumber\\
& = & a (X - X') \cosh (a \tau)~.  \label{91}
\end{eqnarray}
It shows that the redshift drifts with the proper time of the detector. There
is an additional factor $\cosh (a \tau)$ compared to the result in Rindler
coordinates, Eq.(\ref{17}). With time, the redshift would become higher and finally tend to infinity.

Similarly, we can consider light source A on the right of the detector in Figure \ref{Fig1}. The
result shows that a blueshift, namely, $z < 0$,  is observed by the
detector,
\begin{equation}
z_+ = - \dfrac{a (X - X')}{\cosh (a \tau)}~. \label{92}
\end{equation}
In this case, $X < X'$, and the blueshift would become lower with time until it vanishes.

These results, Eqs (\ref{91}) and (\ref{92}), are consistent with  Huang's \cite{huang_uniformly_2008} and those in  the non-relativistic case, qualitatively. Namely, there is 
a redshift drift in the uniformly accelerated reference frames. Further, we
compare all these results in detail, which are summarized in Table
\ref{tab1}. We recover the speed of light, $c$, in the formulations, and set $| X' - X | \equiv L$ for consistency. 
%The results are obtained with an approximation that $| X' - X | \ll \dfrac{a}{c^2}$, while the redshift is exact using the hyperbolic metric.

\begin{table}[!ht]
\begin{center}
	\caption{Redshift $z_- $ and blueshift $z_+$ between co-moving objects in uniformly accelerated reference frames calculated with different approaches. \label{tab1}}
	\begin{tabular*}{80mm}{c@{\extracolsep{\fill}}cc}
		\toprule
		& $z_-  $ & $z_+ $\\
		\hline
		Møller coordinate\cite{landsberg_gravitational_1976} & $\frac{a L}{c^2}$ & $- \frac{a L}{c^2}$\\
		Non-relativity & $\frac{a L}{c^2} \left( 1 - \frac{a   t}{c}
		\right)^{- 2}$ & $- \frac{a   L}{c^2} \left( 1 + \frac{a
			t}{c} \right)^{- 2}$\\
		Huang's\cite{huang_uniformly_2008} & $\dfrac{a L}{c^2} e^{\frac{a \tau}{c}}$ & $- \frac{a L}{c^2} e^{- \frac{a \tau}{c}}$\\
		Hyperbolic metric & $\frac{a L}{c^2} \cosh \left( \frac{a \tau}{c}
		\right)$ & $- \frac{a L}{c^2} \dfrac{1}{\cosh \left( \frac{a
				\tau}{c} \right)}$\\
		\bottomrule
	\end{tabular*}
\end{center}        
\end{table}

\begin{figure}[h]
\begin{center}
\includegraphics[width=1\linewidth]{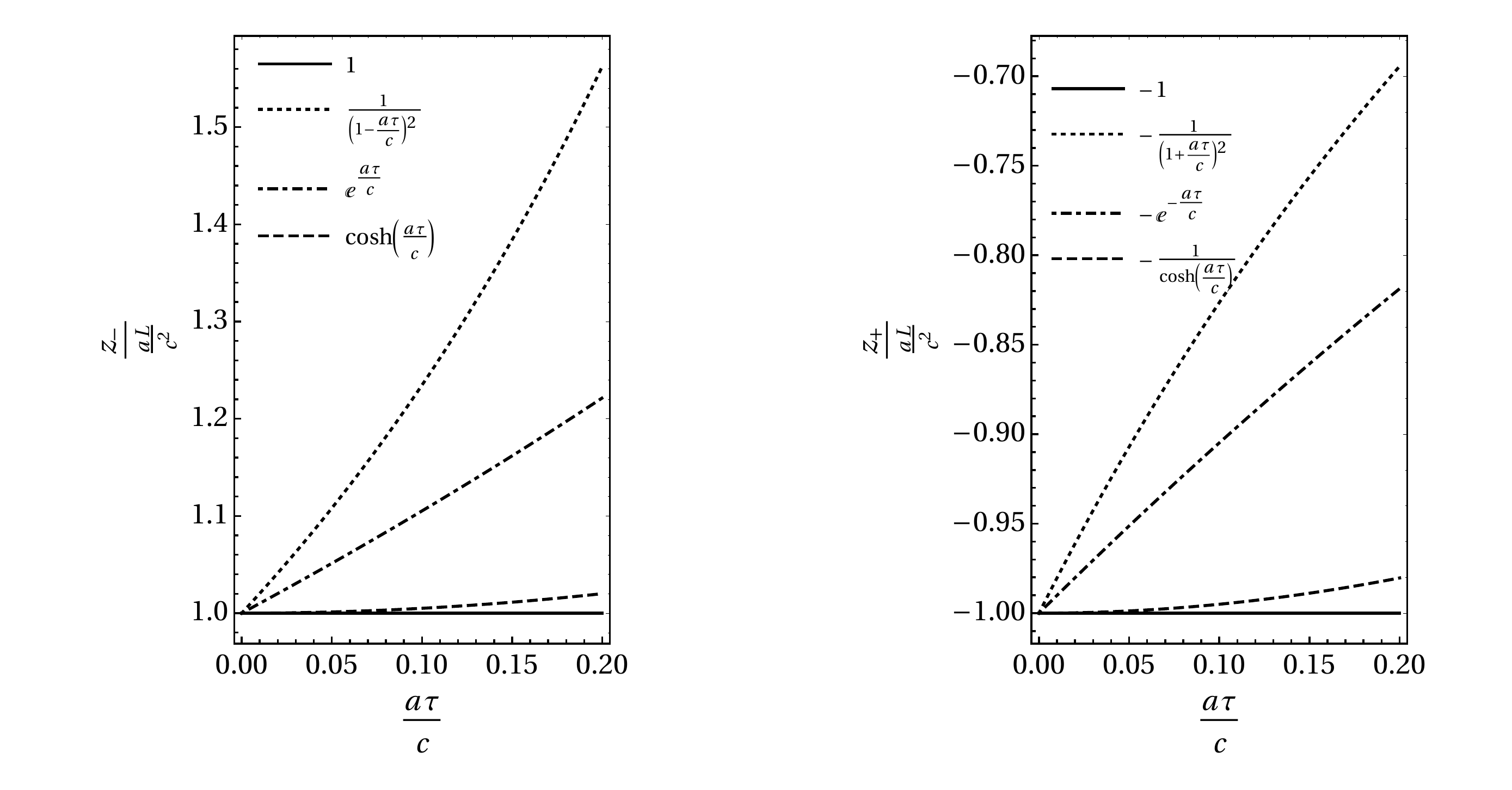}
\caption{Redshift (left panel) and blueshift (right panel) between co-moving objects in uniformly accelerated reference frames as functions of proper time. \label{Fig3}}
\end{center}
\end{figure}

In Figure \ref{Fig3}, the redshift and blueshift  are present as functions of the proper time. In the non-relativistic case, time $t$ is absolute. Therefore, we did not distinguish it with the proper time. The results of Møller
coordinates and the non-relativistic approximation case are contrasting, which are independent of and
sensitive to time, respectively. The redshift calculated with the hyperbolic metric is close to that calculated in Rindler coordinates. In the non-relativity case, it turns to be meaningless when $at\gtrsim c$. In the relativistic case, there is no  limitation, as $a\tau$ is not a three-velocity. 

\subsection{Conformally flat metric and Unruh effect}

\begin{figure}[h]
	{\includegraphics[width=0.6\linewidth]{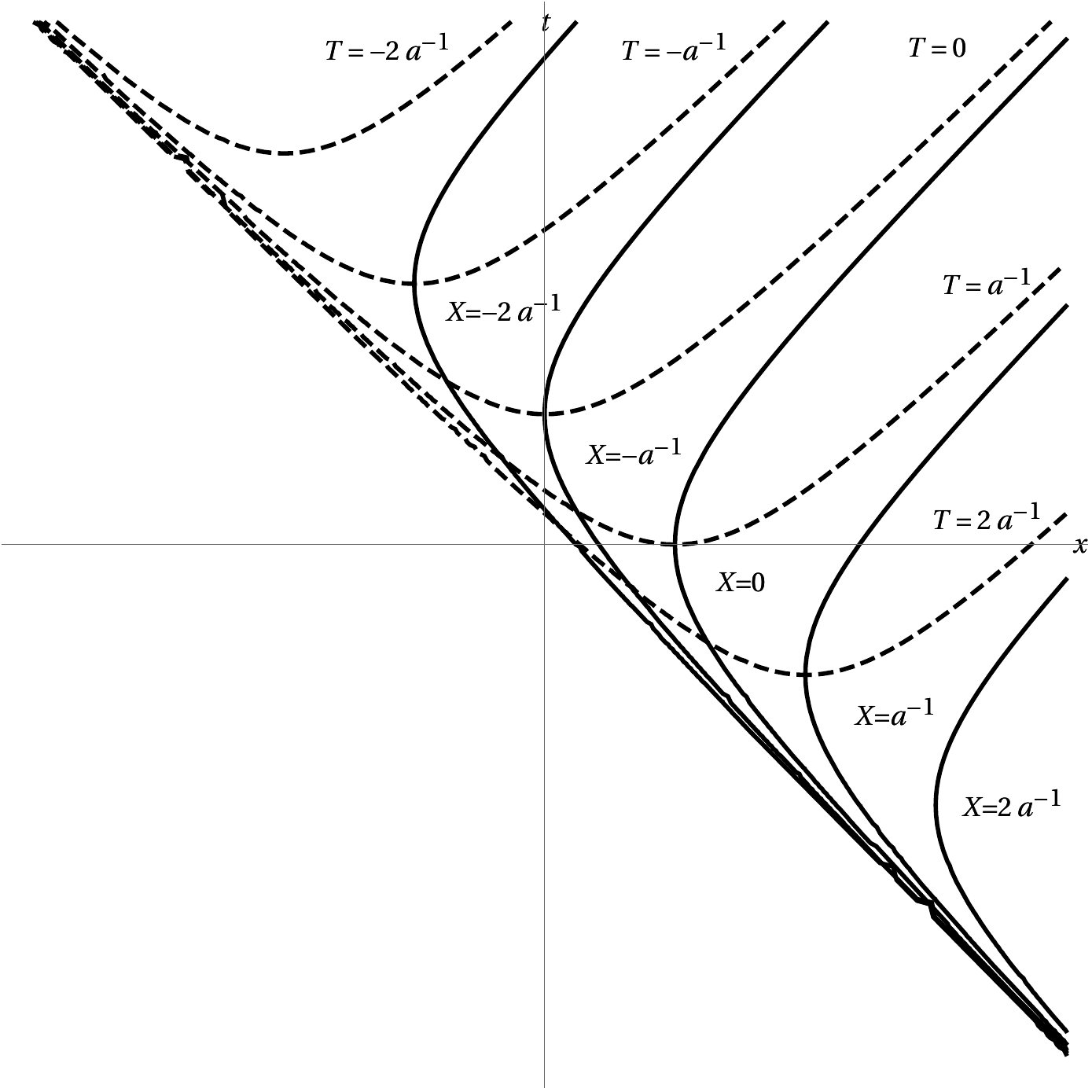}}
	\caption{Coordinate lines of uniformly accelerated frames in  $t$--$x$ diagram. In case of $a>0$, accessible region for accelerated frames is that with $t+x>0$.\label{Fig4}}
\end{figure}

%We expect a conformally flat metric and set $N = \gamma$. 
By making use of Eq.~(\ref{40}) and constraint $N=\gamma$, we get a conformally flat metric,
\begin{equation}
{\rm{d}} s^2 = \dfrac{1}{a^2 (T + X)^2} (-  {\rm{d}} T^2 +  {\rm{d}} X^2) .
\label{93}
\end{equation}
As $N > 0$, it also leads to $- a (T + X) > 0$. Transformation between
the inertial frames and the accelerated frames is given by
\begin{equation}
\left\{\begin{array}{lll}
t & = & \dfrac{1}{2} \left( - \dfrac{1}{a^2 (T + X)} + T - X \right),\\
x & = & \dfrac{1}{2} \left( - \dfrac{1}{a^2 (T + X)} - T + X \right).
\end{array}\right.  \label{94}
\end{equation}
Proper time $\tau$ for co-moving observers can be expressed in terms of the space-time coordinate,
\begin{equation}
\tau = \dfrac{1}{a} \ln \left( - \dfrac{1}{a (T + X)} \right) = \dfrac{1}{a}
\ln (a (t + x)) . \label{95}
\end{equation}
The coordinate lines of the uniformly accelerated frames in $t$--$x$ plane are
presented in Figure \ref{Fig4}.

%Using Eq.~(\ref{95}), we obtain velocity of uniformly a uniformly accelerated particle
%in inertial frame, which can be of the form,
%\begin{equation}
%  \left\{\begin{array}{lllll}
%   u^0 & = & \dfrac{1}{2} \left( \dfrac{1}{a (t + x)} + a (t + x) \right) & = &
%    \cosh (a \tau)\\
%    u^1 & = & - \dfrac{1}{2} \left( \dfrac{1}{a (t + x)} - a (t + x) \right) & =
%    & \sinh (a \tau)
%  \end{array}\right. .
%\end{equation}
%It's also exactly hyperbolic motion. Using coordinate transformation Eq.
%(\ref{94}), the particle is static with respect to the uniformly accelerated
%frame, which is given by,
%\begin{equation}
% \left\{\begin{array}{lllll}
%   u^T & = & - a (T + X) & = & e^{- a \tau}\\
%   u^X & = & 0 &  & 
%  \end{array}\right. .
%\end{equation}

%For a static particle in inertial frame, namely, $\dfrac{ {\rm{d}} t}{ {\rm{d}}
%\tau_0} = 1, \dfrac{ {\rm{d}} x}{ {\rm{d}} \tau_0} = 0$, one can calculate the
%velocity of uniformly accelerated as,
%\begin{eqnarray}
% \upsilon^T & = & N \dfrac{ {\rm{d}} T}{ {\rm{d}} \tau_0} = - \dfrac{1}{2} \left( a
%  (T + X) + \dfrac{1}{a (T + X)} \right) \nonumber\\
%  & = & \cosh (a \tau) 
%\end{eqnarray}
%\begin{eqnarray}
%  \upsilon^X & = & \gamma \tfrac{ {\rm{d}} X}{ {\rm{d}} \tau_0} = \dfrac{1}{2} \left(
%  \dfrac{1}{a (T + X)} - a (T + X) \right) \nonumber\\
%  & = & - \sinh (a \tau) . 
%\end{eqnarray}
%It shows that the inertial motion is hyperbolic motion in the view of
%uniformly accelerated frame.

The Unruh effect states that in uniformly accelerated frames (Rindler coordinates),
the co-moving observers would perceive a thermal distribution of the Minkowski
vacuum \cite{davies_scalar_1975,fulling_radiation_1976,unruh_notes_1976}. The temperature of the distribution is proportional to the constant acceleration
, $a$, in Rindler coordinates. In this subsection, we use
the metric in Eq.~(\ref{93}) to calculate the possible Unruh effect. For
simplicity, we consider a massless Boson.

The Klein--Golden equation for a massless Boson is given by
\begin{equation}
\nabla_{\mu} \nabla^{\mu} \phi = 0,
\end{equation}
where $\nabla_{\mu}$ is the covariant derivative. With our metric, the equation of
motion can be written as
\begin{equation}
% 0 = g^{\mu \nu} \nabla_{\nu} \nabla_{\mu} \phi & = & g^{\mu \nu}   (\partial_{\nu} \partial_{\mu} \phi + \Gamma_{\nu \mu}^{\lambda}  \partial_{\lambda} \phi) = g^{\mu \nu} \partial_{\nu} \partial_{\mu} \phi \nonumber\\
%  & = & \dfrac{1}{a^2 (T + X)^2} (\partial_T^2 \phi + \partial_X^2 \phi), 
g^{\mu \nu} \nabla_{\nu} \nabla_{\mu} \phi = -\dfrac{1}{a^2 (T + X)^2} (\partial_T^2 \phi - \partial_X^2 \phi) = 0~, 
\end{equation}
namely,
\begin{equation}
\partial_T^2 \phi - \partial_X^2 \phi = 0~.
\end{equation}
It is the same as the Klein--Golden equation in a flat space-time. The solution of the
equation can be given by
\begin{equation}
\phi = \dfrac{1}{\sqrt{2 \pi}} \int  {\rm{d}}^2 k \{ \delta (k_0^2 - k^2)
\tilde{\phi} (k_0, k) e^{- i   (k_0 T - k   X)} \}~.
\end{equation}
It can be expanded as
\begin{eqnarray}
\phi & = & \dfrac{1}{\sqrt{2 \pi}} \int_{- \infty}^{\infty} \dfrac{ {\rm{d}} k}{2
	| k |} \{ \tilde{\phi} (e^{- i (| k | T - k   X)} + e^{- i (- | k | T
	- k   X)}) \} \nonumber\\
& = & \dfrac{1}{\sqrt{2 \pi}} \int_0^{\infty} \dfrac{ {\rm{d}} k}{2 k} \{
\tilde{\phi}_k e^{- i   k (T - X)} + \tilde{\phi}_k e^{i   k
  (T + X)}  \nonumber\\
& &  + \tilde{\phi}_{- k} e^{- i   k (T + X)} + \tilde{\phi}_{-
	k} e^{i   k (T - X)} \} \nonumber\\
& = & \phi_- + \phi_+~, 
\end{eqnarray}
where
\begin{equation}
\phi_+ = \dfrac{1}{\sqrt{2 \pi}} \int_0^{\infty} \dfrac{ {\rm{d}} k}{2 k} \{
\tilde{\phi}_{- k} e^{- i   k (T + X)} + \tilde{\phi}_k e^{i  
	k (T +   X)} \} . 
\end{equation}
We focus on left-moving sectors $\phi_+$ of the field. Different
sectors $\phi_-$ and $\phi_+$ would not interact with each other \cite{crispino_unruh_2008}.
We quantize $\phi_+$  in the form of
\begin{eqnarray}
\hat{\phi}_+  =  \dfrac{1}{\sqrt{2 \pi}} \int_0^{\infty} \dfrac{ {\rm{d}}
	k}{\sqrt{2   k}} \{ \hat{b}_k e^{- i   k (T + X)} +
\hat{b}_k^{\dag} e^{i   k (T +   X)} \}, \label{106} 
\end{eqnarray}
where $\hat{b}_k$ is a ladder operator satisfying canonical communication
relations,
\begin{eqnarray}
{}[\hat{b}_k, \hat{b}_{k'}^{\dag}] & = & \delta (k - k')~, \\
\rm{others} & = & 0~, \nonumber
\end{eqnarray}
and
\begin{equation}
\hat{b}_k |0_{\rm{A}} \rangle = 0,
\end{equation}
where $|0_A \rangle$ is a vacuum state in the uniformly accelerated reference frames.
The mode function is read from the field operator in Eq.~(\ref{106}),
\begin{equation}
g_k (T, X) = \dfrac{1}{\sqrt{4 \pi k}} e^{- i   k     (T
	+ X)} . \label{109}
\end{equation}
On the other side, we know the field operator of a left-moving sector in a flat
space-time,
\begin{eqnarray}
\hat{\phi}_+  =  \dfrac{1}{\sqrt{2 \pi}} \int_0^{\infty} \dfrac{ {\rm{d}}
	p}{\sqrt{2     p}} \{ \hat{a}_p e^{- i   p (t + x)} +
\hat{a}_p^{\dag} e^{i   p (t +   x)} \}, 
\end{eqnarray}
where $\hat{a}_p$ is a ladder operator. The canonical communication relations
are given by
\begin{eqnarray}
{}[\hat{a}_p, \hat{a}_{p'}^{\dag}] & = & \delta (p - p')~, \\
\rm{others} & = & 0~. \nonumber
\end{eqnarray}
Moreover, one has
\begin{equation}
\hat{a}_p |0_{\rm{M}} \rangle = 0~,
\end{equation}
where $|0_M \rangle$ is the Minkowski vacuum state. The mode function is of the form,
\begin{equation}
f_p (t, x) = \dfrac{1}{\sqrt{4 \pi p}} e^{- i   p     (t
	+ x)} . \label{113}
\end{equation}
The ladder operators in the accelerated and inertial frames are related to the
so-called Bogolubov transformation, which is given by
\begin{equation}
\left\{\begin{array}{lll}
\hat{a}_p = \int  {\rm{d}} k \{ \alpha_{k   p} \hat{b}_k + \beta_{k
	p}^{\mathord{*}} \hat{b}_k^{\dag} \}~, \\
\hat{b}_k = \int  {\rm{d}} p \{ \alpha_{k   p}^{\mathord{*}} \hat{a}_p -
\beta^{\mathord{*}}_{k   p} \hat{a}_p^{\dag} \}~,
\end{array}\right.  
\end{equation}
where $\alpha_{k   p}$ and $\beta_{k   p}$ are the so-called Bogolubov
coefficients satisfying the relations,
\begin{eqnarray}
\left\{\begin{array}{lll}
\int  {\rm{d}} k \{ \alpha_{k   p} \alpha_{k   p'}^{\mathord{*}} -
\beta_{k   p}^{\mathord{*}} \beta_{k   p'} \} & = & \delta (p -
p')~,\\
\int  {\rm{d}} p \{ \alpha_{k   p}^{\mathord{*}} \alpha_{k  ' p} -
\beta_{k   p}^{\mathord{*}} \beta_{k'   p} \} & = & \delta (k -
k')~,\\
\int  {\rm{d}} k \{ \alpha_{k   p} \beta_{k   p'}^{\mathord{*}} -
\beta_{k   p}^{\mathord{*}} \alpha_{k   p'} \} & = & 0~,\\
\int  {\rm{d}} p \{ \alpha_{k   p}^{\mathord{*}} \beta_{k' p}^{\mathord{*}}
- \beta_{k   p}^{\mathord{*}} \alpha_{k' p}^{\mathord{*}} \} & = & 0~.
\end{array}\right.  
\end{eqnarray}
For the mode functions, orthogonal relations can be derived from the so-called
Klein--Gordon inner product, 
\begin{eqnarray}
(\phi, \chi) & \equiv & i \int_{\Sigma}  {\rm{d}} \Sigma^{\mu} \{
\phi^{\mathord{*}} \nabla_{\mu} \chi - \chi \nabla_{\mu} \phi^{\mathord{*}} \} .
\end{eqnarray}
One can find that
\begin{eqnarray}
\left\{\begin{array}{lll}
(f_p, f_p) & = & \delta (p - p')~,\\
(f_p, f^{\mathord{*}}_p) & = & 0~,\\
(g_k, g_{k'}) & = & \delta (k - k')~,\\
(g_{k  }, g_{k'}^{\mathord{*}}) & = & 0~.
\end{array}\right.  
\end{eqnarray}
In different coordinates, field operator $\hat{\phi}_+$ remains the same
under the Bogolubov transformation. Thus, one can derive the Bogolubov transformation
for the mode functions,
\begin{equation}
g_k = \int  {\rm{d}} p \{ \alpha_{k   p} f_p + \beta_{k   p}
f_p^{\mathord{*}} \} . \label{117}
\end{equation}
From Eq.~(\ref{117}), the Klein--Gordon inner product can be used to calculate
the Bogolubov coefficients,
\begin{eqnarray}
\left\{\begin{array}{lll}
\alpha_{k   p} & = & (f_p, g_k),  \label{118}\\
\beta_{k   p} & = & - (f^{\mathord{*}}_k, g_k) .  \label{119}
\end{array}\right.
\end{eqnarray}
What the accelerated observers perceive in the Minkowski vacuum is formulated as the expectation value of occupation number operators $N_k$ of the accelerated observers for the Minkowski vacuum state,
\begin{eqnarray}
\langle 0_{\rm{M}} | N_k |0_{\rm{M}} \rangle & = & \langle 0_{\rm{M}} |
\hat{b}_k^{\dag} \hat{b}_k |0_{\rm{M}} \rangle 
%= \langle 0_{\rm{M}} | \int  {\rm{d}} 
%  p \{ \alpha_{k   p} \hat{a}_p^{\dag} - \beta_{k   p} \hat{a}_p
%  \} \int  {\rm{d}} p' \{ \alpha_{k   p'}^{\mathord{*}} \hat{a}_{p'} -
%  \beta^{\mathord{*}}_{k   p'} \hat{a}_{p'}^{\dag} \} |0_{\rm{M}} \rangle
%  \nonumber\\
= \int  {\rm{d}} p \beta_{p   k} \beta_{p   k}^{\mathord{*}}~.
\end{eqnarray}
It shows that the expectation value only involves the Bogolubov coefficients, $\beta_{p   k}$.

As in Refs. \cite{takagi_vacuum_1986,crispino_unruh_2008}, we use light-cone coordinates to
calculate the Bogolubov coefficients from Eq.~(\ref{119}) for a
given null hypersurface. The null coordinates are usually closely related to radiation \cite{bondi_gravitational_1960,bondi_gravitational_1962}. Light-cone coordinates of the inertial frames and the uniformly accelerated frames are given by $(u,v)=(t-x, t+x)$ and $(U,V)=(T-X,T+X)$, respectively.
% Light-cone coordinate of inertial frame $(t, x)$ and uniformly accelerated frame $(T, X)$ are given by
%\begin{equation}
%  \left\{\begin{array}{lll}
%    u & = & t - x,\\
%    \upsilon & = & t + x.
% \end{array}\right. \label{121}
%\end{equation}
%and
%\begin{equation}
%  \left\{\begin{array}{lll}
%    U & = & T - X,\\
%    V & = & T + X.
%  \end{array}\right. \label{122}
%\end{equation}
From Eq.~(\ref{94}) and the light-cone coordinates, %(\ref{121}) and (\ref{122}), 
the transformation between the uniformly accelerated frames
and the inertial frames is obtained,
\begin{equation}
\left\{\begin{array}{lll}
u & = & U,\\
\upsilon & = & - \dfrac{1}{a^2 V}.
\end{array}\right. 
\end{equation}
In the light-cone coordinates, the metric of the accelerated frames can be
rewritten as
\begin{equation}
{\rm{d}} s^2 = -  {\rm{d}} u  {\rm{d}} \upsilon = \dfrac{1}{a^2 V^2}  {\rm{d}} U  {\rm{d}}
V.
\end{equation}
We choose the null hypersurface as
\begin{equation}
\Phi (U, V) \equiv U = \rm{cosntant} .
\end{equation}
One can find that it is the only non-trivial null hypersurface for calculating
the possible Unruh effect, otherwise it would lead to $\beta_{p   k}
\equiv 0$. The normal vectors of the null hypersurface are null vectors, $\xi_{\nu} = - \partial_{\nu} \Phi=-\delta^0_\nu$.
%\begin{equation}
%  \xi_{\nu} = - \partial_{\nu} \Phi = \left(\begin{array}{c}
%    - 1\\
%    0
%  \end{array}\right) \label{126},
%\end{equation}
%and
%\begin{equation}
% \xi^{\mu} = \left(\begin{array}{c}
%  \dfrac{\partial U}{\partial \lambda}\\
%   \dfrac{\partial V}{\partial \lambda}
%  \end{array}\right), \label{127}
%\end{equation}
We use $\lambda$ to parametrize the integral curve $(U (\lambda), V (\lambda))$
of $\xi^{\mu} (\lambda)$. As the null vector , $\xi^{\mu}$, is also tangent to
the null hypersurface, $\lambda$ also can be used to
parametrize the null hypersurface in the two-dimensional case. %From Eqs.~(\ref{126}) and (\ref{127}), 
Thus, we can solve the integral curve, $(U, V)$, as
\begin{equation}
\left\{\begin{array}{lll}
U & = & \rm{constant},\\
V & = & \dfrac{1}{2 a^2 \lambda}.
\end{array}\right.  \label{128}
\end{equation}
Using Eq.~(\ref{128}), we obtain the volume elements of the null hypersurface \cite{poisson_relativists_2004},
\begin{equation}
{\rm{d}} \Sigma \equiv \epsilon_{\mu \nu} \zeta^{\mu} \xi^{\nu}  {\rm{d}} \lambda
= \dfrac{1}{2 a^2 V^2}  {\rm{d}} V, \label{129}
\end{equation}
where $\epsilon_{\mu \nu}$ is the Levi-Civita tensor and $\zeta^{\mu}$ is an
auxiliary null vector satisfying $\xi^{\mu} \zeta_{\mu} = - 1$ and
$\zeta^{\mu} \zeta_{\mu} = 0$. From Eq.~(\ref{129}), the directed surface
element is obtained,
\begin{equation}
{\rm{d}} \Sigma^{\mu} = - \xi^{\mu}  {\rm{d}} \Sigma = \delta^{\mu}_1  {\rm{d}} V.
\label{130}
\end{equation}
Here, we choose that $a > 0$. For the metric with $N > 0$, it leads to $V <
0$.

Form Eqs.~(\ref{109}), (\ref{113}), (\ref{118}), and (\ref{130}),
we can calculate the Bogolubov coefficients,
%\end{multicols}
\begin{eqnarray}
\alpha_{k   p} & = & (f_p, g_k) \nonumber\\
& = & i \int_{- \infty}^0  {\rm{d}} V \left\{ \dfrac{1}{\sqrt{4 \pi p}} e^{i
	p   \upsilon} \partial_V \dfrac{1}{\sqrt{4 \pi k}} e^{- i
	k   V} - \dfrac{1}{\sqrt{4 \pi k}} e^{- i   k  
	V} \partial_V \dfrac{1}{\sqrt{4 \pi p}} e^{i   p   \upsilon}
\right\} \nonumber\\
& = & \dfrac{1}{4 \pi \sqrt{p   k}} \left( k \int_0^{\infty}  {\rm{d}} V
e^{i \left( k   V + \frac{p}{a^2 V} \right)} - \dfrac{p}{a^2}
\int_0^{\infty}  {\rm{d}} \left( \dfrac{1}{V} \right)   e^{i \left( k
	V + \frac{p}{a^2 V} \right)} \right) \nonumber\\
& = & \dfrac{1}{\pi a} \int_1^{\infty}  {\rm{d}} \sqrt{\eta^2 - 1} \left\{ e^{2
	\sqrt{\frac{p   k}{a^2}} i   \eta  } \right\}, 
\end{eqnarray}
and
\begin{eqnarray}
\beta_{k   p} & = & - (f^{\mathord{*}}_p, g_k) \nonumber\\
& = & - i \int_{- \infty}^0  {\rm{d}} V \left\{ \dfrac{1}{\sqrt{4 \pi p}} e^{-
	i   p   \upsilon} \partial_V \dfrac{1}{\sqrt{4 \pi k}} e^{- i
	k   V} - \dfrac{1}{\sqrt{4 \pi k}} e^{- i   k  
	V} \partial_V \dfrac{1}{\sqrt{4 \pi p}} e^{- i   p   \upsilon}
\right\} \nonumber\\
& = & - \dfrac{1}{4 \pi \sqrt{p   k}} \left( \dfrac{p}{a^2}
\int_0^{\infty}  {\rm{d}} \left( \dfrac{1}{V} \right) e^{i \left( k   V -
	\frac{p}{a^2 V} \right)} + k \int_0^{\infty}  {\rm{d}} V   e^{i \left( k
	V - \frac{p}{a^2 V} \right)} \right) \nonumber\\
& = & - \dfrac{1}{2 \pi a} \int_{- \infty}^{\infty}  {\rm{d}} \sqrt{\eta^2 + 1}
\left\{ e^{2 \sqrt{\frac{k   p}{a^2}} i   \eta} \right\} . 
\end{eqnarray}
%\begin{multicols}{2}
With the tricks of $\eta \rightarrow e^{i \epsilon} \infty$ and $\epsilon
> 0$ \cite{crispino_unruh_2008}, the Bogolubov coefficients can be obtained explicitly,
\begin{eqnarray}
\alpha_{k   p} & = & \dfrac{1}{\pi a} K_1 \left( - 2 i \sqrt{\dfrac{p
		k}{a^2}} \right), 
\end{eqnarray}
and
\begin{eqnarray}
\beta_{k   p} & = & - \dfrac{1}{2 \pi a} \left( 1 + i   K_1
\left( 2 \sqrt{\dfrac{p   k}{a^2}} \right) \right. \nonumber \\
& & \left. + \dfrac{\pi}{2} \left( L_1
\left( 2 \sqrt{\dfrac{p   k}{a^2}} \right) - I_1 \left( 2
\sqrt{\dfrac{p   k}{a^2}} \right) \right) \right. \nonumber \\
& & + \left. \dfrac{1}{2 \pi} G^{1, 3}_{3,  1} \left( \dfrac{a^4}{k  ^2 p^2}, 2 \middle| \begin{array}{ccc}
\dfrac{1}{2} & 1 & \dfrac{3}{2}\\
& 1 & 
\end{array} \right) \right), 
\end{eqnarray}
where $K_1$ and $I_1$ are the modified Bessel functions of the second and first
kind, respectively, $L_1$ is the modified Struve function, and $G^{1, 3}_{3, 1}$
is the generalized Meijer G function. The above shows that the Bogolubov coefficients,
$\alpha_{k   p}$ and $\beta_{k   p}$, are completely different from
those in Rindler coordinates \cite{davies_scalar_1975,takagi_vacuum_1986,crispino_unruh_2008}. This suggests
that the expectation value  might not be the form of that calculated in Rindler
coordinates. We finally obtain the expectation value in our uniformly accelerated frames as follows:
%\end{multicols}
%\ruleup
\begin{eqnarray}
\langle 0_{\rm{M}} | \hat{N}_k |0_{\rm{M}} \rangle & = & \int  {\rm{d}} p
\beta_{  k   p} \beta_{  k   p}^{\mathord{*}}
\nonumber\\
& = & \int_0^{\infty}  {\rm{d}} p \left\{ \left( - \dfrac{1}{2 \pi a} \int_{-
	\infty}^{\infty}  {\rm{d}} \sqrt{\eta^2 + 1} \left\{ e^{2 \sqrt{\frac{k
			p}{a^2}} i   \eta} \right\} \right) \left( - \dfrac{1}{2 \pi
	a} \int_{- \infty}^{\infty}  {\rm{d}} \sqrt{\eta^{\prime 2} + 1} \left\{ e^{2
	\sqrt{\frac{k   p}{a^2}} i   \eta'} \right\}
\right)^{\mathord{*}} \right\} \nonumber\\
& = & \dfrac{1}{8 \pi^2 k} \int_0^{\infty}  {\rm{d}} \chi \int_{-
	\infty}^{\infty}  {\rm{d}} \eta \int_{- \infty}^{\infty}  {\rm{d}} \eta' \left\{
\dfrac{\chi   \eta   \eta'}{\sqrt{\eta^2 + 1}
	\sqrt{\eta^{\prime 2} + 1}} e^{i   z (\eta - \eta')} \right\}
\nonumber\\
& \equiv & \dfrac{\Lambda}{k} \label{135}, 
\end{eqnarray}
where $\chi = 2 \sqrt{\dfrac{k   p}{a^2}}$, and
\begin{eqnarray}
\Lambda & \equiv & \dfrac{1}{8 \pi^2} \int_0^{\infty}  {\rm{d}} \chi \int_{-
	\infty}^{\infty}  {\rm{d}} \eta \int_{- \infty}^{\infty}  {\rm{d}} \eta' \left\{
\dfrac{\chi \eta   \eta'}{\sqrt{\eta^2 + 1} \sqrt{\eta^{\prime 2} +
		1}} e^{i   z (\eta - \eta')} \right\} \nonumber\\
& < & \dfrac{1}{8 \pi^2} \int_0^{\infty} \int_{- \infty}^{\infty} \int_{-
	\infty}^{\infty}  {\rm{d}} \chi  {\rm{d}} \eta  {\rm{d}} \eta' \left\{ \dfrac{\chi
	\eta \eta'}{| \eta | | \eta' |} e^{i   \chi (\eta - \eta')} \right\}  =  \left. \dfrac{1}{4 \pi^2} \ln   \chi \right|^{\infty}_0 . 
\end{eqnarray}
%\begin{multicols}{2}
Constant $\Lambda$ is divergent. We use the tricks of $\eta \rightarrow
e^{i \epsilon} \infty$ to obtain the last equal sign.

Firstly, this shows that the distribution of Eq.(\ref{135}) is independent of the
acceleration for our accelerated frame.  Secondly,  the expectation value of the number operator  for the Minkowski vacuum state is a non-thermal distribution of
$k$. The uniformly accelerated observers can not perceive temperature in the
Minkowski vacuum.  %It might suggest that the distribution might be not physical.

\section{Conclusions and discussions }\label{VI}

In this study, we constructed new uniformly accelerated reference frames based on the 3+1 formalism in adapted coordinates for uniformly accelerated observers $u$. The norms of four-accelerations of co-moving observers are the same in the uniformly accelerated reference frames. The inertial motion in the view of a uniformly accelerated observer can be formulated as a uniformly accelerated motion. % in the view of the accelerated frames. 
Moreover, the space-time would be deformed by a non-inertial effect. We also presented explicit metrics and coordinate transformations. In contrast with Rindler coordinates, the redshift for co-moving observers would drift with time, in our accelerated frames. This is consistent with earlier results \cite{huang_uniformly_2008}. Besides, we calculated the possible Unruh effect and showed a non-thermal distribution for the Minkowski vacuum state perceived by the uniformly accelerated observers.

From our approaches, the constraint equations in Eq.~(\ref{40}) for uniformly accelerated
frames are under-determined. It results in that the metric of the reference frames is
non-unique. The degrees of freedom for different measurements and synchronisation
conventions are allowed. Firstly, Eq.~(\ref{40}) is invariant under
transformation $T \rightarrow f (T)$ and $X \rightarrow g (X)$. This is owing to integrating factor $N$ and $\gamma$ being not unique. It indicates
that the time and distance can be measured with different clocks and rulers.
Secondly, there are different simultaneous hypersurfaces for the explicit
solutions in Section \ref{IV}. It suggests different synchronisation conventions. Moreover, it is rather interesting to explore what kinds of conventions are physically operational.

After Huang \cite{huang_uniformly_2008} first suggested that a redshift drift can be observed in uniformly accelerated reference frames, we also provided a similar prediction with different approaches. The redshift of a co-moving object is different from that in Rindler frames, because the co-moving objects of these reference frames are defined differently. As mentioned in Ref.~\cite{Li_2014}, the concept of relative velocities between non-inertial observers is usually ambiguous. Thus, it is non-trival to obtain a convention for co-moving objects via introducing  physical reference frames. Moreover, if a set-up of co-moving objects can be realized in experiment as shown in Figure~\ref{Fig1}, we might expect that these results derived from different frames can be tested in the future.

\acknowledgments{The authors wish to thank Prof.~Chao-Guang Huang for useful discussions.}

\vspace{10mm}

\bibliography{citation}
%\end{multicols}
\end{document}